\documentclass[conference]{IEEEtran}
\IEEEoverridecommandlockouts                              

\usepackage{graphics} 
\usepackage{epsfig} 
\usepackage{comment}
\usepackage{url}
\usepackage{cite}
\usepackage{balance} 
\usepackage{amsmath} 
\usepackage{amssymb}  
\usepackage{amsfonts}
\usepackage{type1cm}
\usepackage{multirow}
\usepackage{textcomp}

\usepackage[whole]{bxcjkjatype} 

\def\BibTeX{{\rm B\kern-.05em{\sc i\kern-.025em b}\kern-.08em
    T\kern-.1667em\lower.7ex\hbox{E}\kern-.125emX}}
\begin{document}



\title{Brain-inspired probabilistic generative model \\for double articulation analysis of spoken language
    \thanks{
        *This work was partially supported by MEXT/JSPS KAKENHI under Grant numbers JP16H06569, JP17H06315, JP21H04904.
    }
}

\author{

\IEEEauthorblockN{Akira Taniguchi}
\IEEEauthorblockA{\textit{College of Information Science and Engineering} \\
\textit{Ritsumeikan University}\\
Kusatsu, Shiga, Japan\\
a.taniguchi@em.ci.ritsumei.ac.jp}
\and
\IEEEauthorblockN{Maoko Muro}
\IEEEauthorblockA{\textit{Graduate School of Information Science and Engineering} \\
\textit{Ritsumeikan University}\\
Kusatsu, Shiga, Japan\\
muro.maoko@em.ci.ritsumei.ac.jp}
\and
\IEEEauthorblockN{Hiroshi Yamakawa}
\IEEEauthorblockA{\textit{The Whole Brain Architecture Initiative} \\
\textit{/ The University of Tokyo / RIKEN} \\
Taito-ku, Tokyo, Japan\\
ymkw@wba-initiative.org}
\and
\IEEEauthorblockN{Tadahiro Taniguchi}
\IEEEauthorblockA{\textit{College of Information Science and Engineering} \\
\textit{Ritsumeikan University}\\
Kusatsu, Shiga, Japan\\
taniguchi@em.ci.ritsumei.ac.jp}
}

\maketitle

\begin{abstract}

The human brain, among its several functions, analyzes the double articulation structure in spoken language, i.e., \textit{double articulation analysis} (DAA).
A hierarchical structure in which words are connected to form a sentence and words are composed of phonemes or syllables is called a double articulation structure.
Where and how DAA is performed in the human brain has not been established, although some insights have been obtained. 
In addition, existing computational models based on a probabilistic generative model (PGM) do not incorporate neuroscientific findings, and their consistency with the brain has not been previously discussed.
This study compared, mapped, and integrated these existing computational models with neuroscientific findings to bridge this gap, and the findings are relevant for future applications and further research.
This study proposes a PGM for a DAA hypothesis that can be realized in the brain based on the outcomes of several neuroscientific surveys. 
The study involved 
(i) investigation and organization of anatomical structures related to spoken language processing, 
and 
(ii) design of a PGM that matches the anatomy and functions of the region of interest.
Therefore, this study provides novel insights that will be foundational to further exploring DAA in the brain. 

\end{abstract}



\section{Introduction}
\label{sec:introduction}

%
Referencing the brain allows for accelerated development of cognitive architectures by reducing the design space~\cite{yamakawa2021whole}.
The brains of humans and other animals have various real-world functions.
Brain-inspired cognitive architectures are useful for building intelligent robots that can coexist with humans and operate in the real world.
Similar research areas include biologically-inspired cognitive architectures~\cite{Samsonovich2016-tn, Goertzel2010-hy} and cognitive computational neuroscience~\cite{ Kriegeskorte2018-xw}. 
This study focused on the function of the \textit{double articulation analysis} (DAA) of spoken language in the human brain. 
DAA involves finding hierarchical structures such as phonemes and words in spoken language. 
Building a computational model that mimics the human brain's spoken language processing becomes a bridge to understanding and supporting the development of communication robots.

%
%
There are gaps in neuro/cognitive science findings and artificial intelligence techniques for DAA of spoken language; however, it is important to integrate them.
It is assumed that some regions in the brain are responsible for the processing of the double articulation structure.
There is a near consensus on the regional partitioning of the auditory cortex in non-human primates~\cite{Kayser2007-ee}. 
However, where and how DAA is performed in the human brain has not been established, although some insights into this process have been obtained\cite{Bhaya-Grossman2022-gg}.
%
%
In contrast, there are computational models that allow for DAA~\cite{taniguchi2016nonparametric,okuda2021double,Taniguchi2022UnsupervisedMW}.
The nonparametric Bayesian double articulation analyzer (NPB-DAA)~\cite{taniguchi2016nonparametric} is an unsupervised learning method based on Bayesian inference in a probabilistic generative model (PGM) that assumes that time-series data has a double-articulation structure.
However, these computational models do not incorporate neuroscientific findings, and their correspondence to the brain has not been discussed.
This study compared, mapped, and integrated these existing computational models with neuroscientific findings.

%
This study is part of the grand challenge of realizing a whole-brain architecture using PGM~\cite{taniguchi2021whole}. 
Related approaches have modeled the biologically-inspired function in whole-brain format~\cite{Eliasmith1202, sagar2016}; however, they have not been studied in a manner that can be implemented in a real robot.
The reason for modeling a brain-inspired architecture using PGMs is that the Neuro-SERKET (\textit{symbol emergence in the robotics tool kit}) is available for both theoretical and practical use~\cite{Taniguchi2020neuro}. 
Whole-brain PGM is expected to integrate with PGM sub-modules for multiple brain regions using Neuro-SERKET. 
Therefore, the PGM for DAA proposed in this study can be used as a module to represent an integrated whole-brain cognitive architecture.

%
Brain reference architecture (BRA) can be developed using the \textit{structure-constrained interface decomposition} (SCID) method~\cite{yamakawa2021whole}. 
The SCID method creates a brain information flow (BIF), which is a data flow diagram that shows the information processing architecture of the brain while defining the top-level function (TLF) and the region of interest (ROI). 
This allows the construction of a hypothetical component diagram (HCD) that systematizes the TLFs for which the ROIs are responsible so that they are consistent with the anatomical structure. 
Therefore, the SCID method can be used to construct a computational model in terms of neuroscience and engineering. 
This study used the SCID method to construct a PGM for DAA by generating hypotheses based on anatomical and neuroscientific findings of the brain.
%
%
The contributions of this study are as follows:
\begin{itemize}
  \item[(i)] Investigation of the anatomical structure and function of the brain regions involved in the DAA of spoken language and their compilation into a brain information flow diagram.
  \item[(ii)] Construction of a novel PGM consistent with the anatomical structure and function of the brain based on existing computational models of DAA.
\end{itemize}

\section{Preliminaries}
This section describes the cognitive and developmental studies related to lexical acquisition by infants and the existing PGM-based computational models for DAA.

\subsection{Lexical acquisition by infants}
Infants can discover a lexicon from minimal prior knowledge based on the statistical regularities of speech stimuli and the co-occurrence of other sensory stimuli. 
Saffran et al. found that (i) distributional, (ii) prosodic, and (iii) co-occurrence cues are important for lexical acquisition in infants~\cite{saffran1996word}. 
Distributional cues are statistical regularities related to the phonological features of an utterance, and prosodic cues are the intonation and the silent intervals contained in utterances.
Co-occurrence cues are provided by sensory stimuli that coincide with a particular utterance. Supplemental/top-down information such as context can be used to improve segmentation accuracy in noisy environments.

\subsection{Existing computational models for Double Articulation Analysis}
This study focused on phoneme and word discovery using phonological~\cite{taniguchi2016nonparametric}, prosodic~\cite{okuda2021double}, and co-occurrence cues of speech signals~\cite{Taniguchi2022UnsupervisedMW}.
These methods provide constructive approaches to infant lexical acquisition~\cite{saffran1996word}. 
However, they do not take neuroscientific findings into account.

\textbf{HDP-HLM}: 
NPB-DAA is a method for word and phoneme acquisition that evaluates language and acoustic models through unsupervised learning~\cite{taniguchi2016nonparametric}. 
The parameter inference method applies blocked Gibbs sampling to the \textit{hierarchical Dirichlet process hidden language model} (HDP-HLM). 
HDP-HLM is a PGM for time-series data that extends the hierarchical Dirichlet process hidden semi-Markov model (HDP-HSMM)~\cite{johnson2013}. 
The language model represents the transition probabilities between words, and the acoustic model represents the relationship between each phoneme and acoustic features. 
Acoustic features are obtained by dimensional compression of the speech spectrum by conversion to the Mel Frequency Cepstral Coefficient (MFCC) or deep sparse autoencoder with parametric bias in the hidden layer (DSAE-PBHL)~\cite{nakashima2019unsupervised}.

\textbf{Prosodic HDP-HLM}: 
Okuda et al. proposed a PGM that extends HDP-HLM to utilize prosodic features such as fundamental frequency and silent intervals for lexical acquisition from natural speech signals that include prosody~\cite{okuda2021double}.
Prosody affects the duration distribution, which represents the time length of phonemes in speech. In a lexical acquisition experiment involving spoken utterances, the use of prosodic features was shown to outperform phonological-only speech unit discovery methods~\cite{Taniguchi2022UnsupervisedMW}.

\textbf{HDP-HLM+MLDA}: 
Murakami et al. proposed a PGM for discovering speech units using distributional and co-occurrence cues~\cite{Taniguchi2022UnsupervisedMW}.
HDP-HLM+MLDA is a model that integrates HDP-HLM and multimodal latent Dirichlet allocation (MLDA)~\cite{nakamura2011grounding}.
MLDA is a multimodal extension of the latent Dirichlet allocation~\cite{blei2003latent} and is capable of forming object categories from robot sensor observations. 
This method acquires words and phonemes from speech signals through unsupervised learning, as well as utilizes object categories based on multiple modalities such as vision, haptics, and hearing. 
MLDA-segmented words (equivalent to nouns and adjectives) describe object features with accuracy.

\section{Approach: BRA construction}
This section describes the methodology for constructing PGM as a BRA.
Based on this methodology, a brain interneuron site responsible for processing related to DAA was hypothesized.

\begin{figure}[tb]
  \begin{center}
    \includegraphics[width=\linewidth]{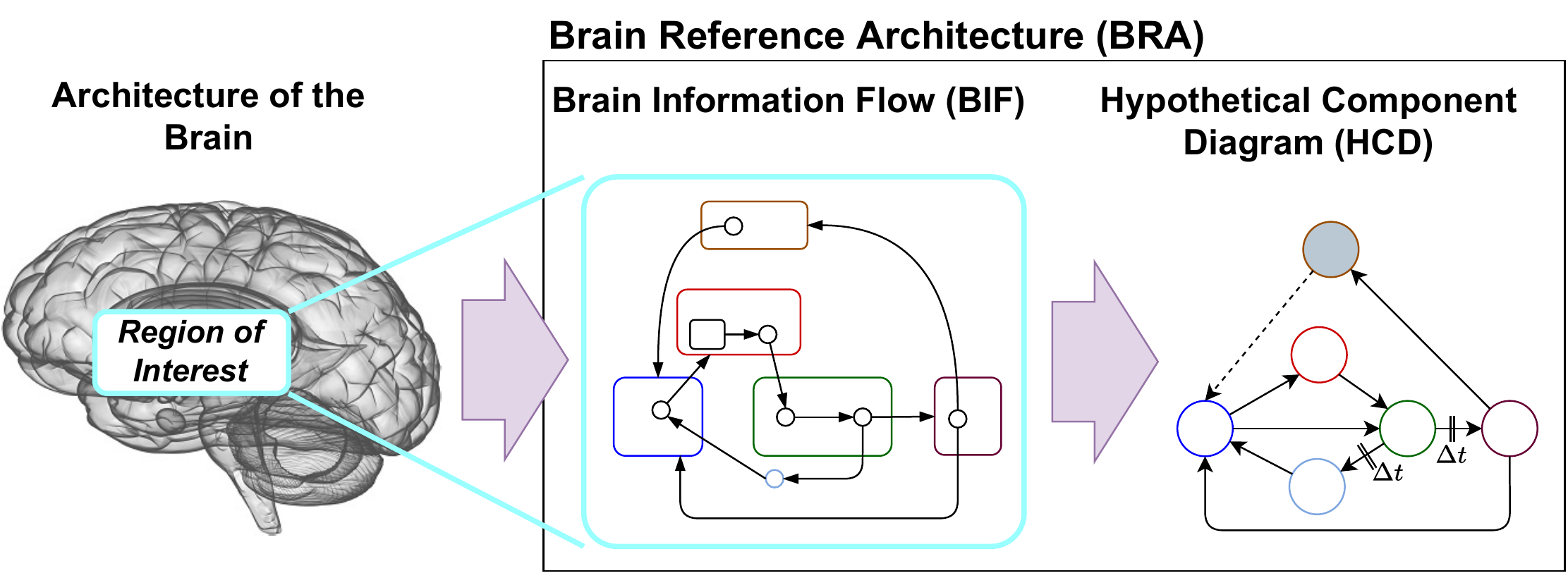}
    \caption{Modeling flow for the BRA and hypothetical component diagram (HCD).
    In this study, HCD is represented as a graphical model representation in the probabilistic generative model (PGM).}
    \label{BRA_PGM}
  \end{center}
\end{figure}

\subsection{Brain reference architecture (BRA)}
This study adopted a brain-referenced architecture-driven developmental approach that builds modules for whole-brain architecture development~\cite{yamakawa2021whole}. 
This is an architecture-oriented approach to developing brain-based machine learning models. 
By describing and publishing the BRA in a unified format, other researchers can refer to it to support their studies.

%
Figure~\ref{BRA_PGM} shows the relationship between the components of the model and information processing in the brain. 
The brain information flow (BIF) associated with anatomical structures is a directed graph consisting of subcircuits and connections that represent the anatomical structure of neural circuits in the brain. 
In principle, the IDs of brain regions on the BIF conform to the \textit{Allen Developing Human Brain Atlas Ontology} (DHBA)~\cite{Ding2017-lc}. 
In this study, the IDs of the DHBA were used as much as possible.
The hypothetical component diagram (HCD), related to the computational functions of BIF, is a directed graph describing the dependencies of its components and can be associated with any BIF subgraph.
The functional mechanisms described by HCD are hypotheses, and neuroscientific findings should offer hypotheses from various perspectives.
It is useful to consider HCD candidates that cannot be unambiguously excluded based on neuroscientific findings.

\subsection{Structure-constrained interface decomposition (SCID)} 

The SCID method provides information on computational functions useful for software implementation, which is mainly constrained by anatomical structures, and enables the construction of functional hypotheses for several brain regions.
This is necessary to ensure consistency between neuroscientific findings and engineering feasibility and to build brain-inspired models.
The architecture consists primarily of mesoscopic-level anatomical data of the brain and functional mechanism data consistent with that knowledge.
Using the SCID method, Fukawa et al.~\cite{fukawa2020identifying} first applied the concept to the neural circuitry of the hippocampal formation and medial entorhinal cortex, and the circuit performing path-integration functions was identified. 
Subsequently, the SCID method was formally adopted by Yamakawa~\cite{yamakawa2020revealing}.

%
With the SCID method, it is possible to obtain an HCD that matches the anatomical structure in the BIF of the ROI mainly by the following three-step process:
\begin{itemize}
    \item[1.] BIF construction: investigation of anatomical knowledge in the ROI.
    \item[2.] Consistent determination of the ROI and TLF (including creation of provisional component diagram).
    \item[3.] HCD creation.\\
        \textbf{Step 3-A}: Enumeration of candidate component diagrams (exhaustive of possibilities).\\
        \textbf{Step 3-B}: Rejection of HCDs that are inconsistent with various scientific findings.
\end{itemize}
Steps 1 and 2 are performed in relation to each other.

%
This study used the SCID method to design an HCD that is consistent with the anatomical structures in the superior temporal gyrus and the temporal frontal lobes, including the primary auditory cortex in the left hemisphere. 
From the above, one HCD was created by applying steps 1 and 2 to the DAA. 
This study deals with PGMs as the corresponding form of HCD. 
The task of mapping the link structure of the PGM to the structure of the actual neural circuit is called \textit{generation-inference process allocation}~\cite{taniguchi2021hippocampal}.

\section{Neuroscientific findings related to spoken language processing}
This section summarizes the survey conducted to determine the ROI and TLF.
First, the ROI and TLF of this study were described using a BIF that summarized the content of the survey. Next, the survey findings were divided into anatomical and functional. By integrating these surveys, the ROI responsible for DAA was determined.

\begin{figure}[!tb]
  \begin{center}
    \includegraphics[width=0.9\linewidth]{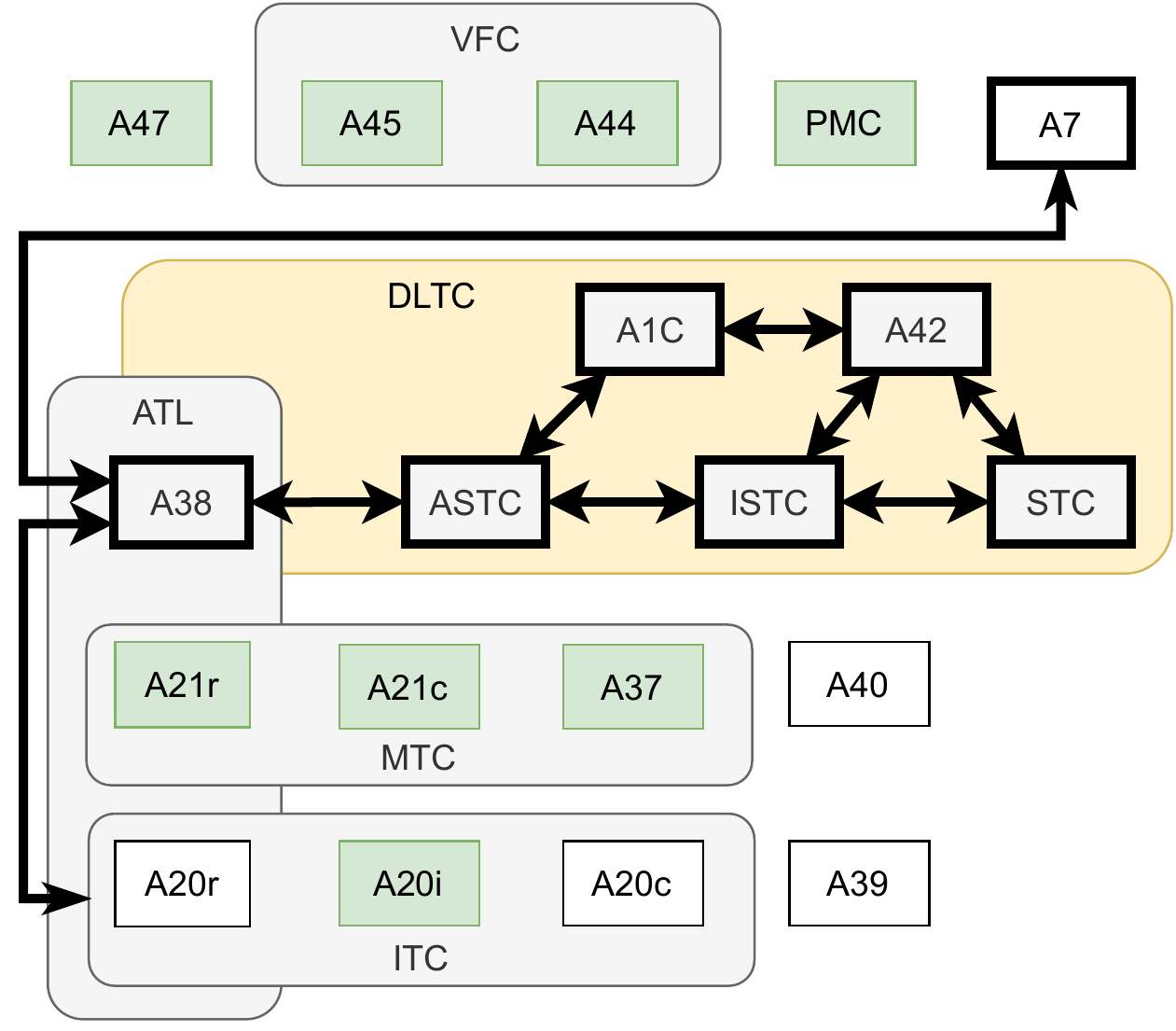}
    \caption{Language-related brain networks as initial scope of investigation~\cite{Bhaya-Grossman2022-gg, Hamilton2021-fc, Scott2017-im,Jasmin2019-wb,Markov2014-ly,Fan2016-wp, Pascual2015-ju}:
    %
    The correspondence of the region names is shown as follows:
    primary auditory cortex (A1C),
    secodary auditory cortex (A42),
    anterior (rostral) superior temporal cortex (ASTC),
    intermediate superior temporal cortex (ISTC),
    posterior (caudal) superior temporal cortex (STC),
    temporal polar cortex (A38),
    %
    lateral orbital frontal cortex (A47),
    rostral portion of ventrolateral prefrontal cortex (A45), 
    caudal portion of ventrolateral prefrontal cortex (A44), 
    premotor cortex (PMC), 
    %
    caudal division of posterodorsal (superior) parietal cortex (A7),
    rostral subdivision of area 21 (A21r), 
    caudal subdivision of area 21 (A21c), 
    lateral temporal-occipital cortex (A37),
    rostral subdivision of area 20 (A20r),
    intermediate subdivision of area 20 (A20i),
    caudal subdivision of area 20 (A20c),
    rostral division of posteroventral inferior parietal cortex (A40), 
    caudal division of posteroventral inferior parietal cortex (A39).
    %
    }
    \label{fig:LanguageRelatedBrainNetwork}
  \end{center}
\end{figure}

\subsection{Top-level function and region of interest}
The TLF in this study was a DAA of speech signals in the brain, where the input was a speech signal consisting of sentences with multiple connected words, and the output was a segmented word sequence.
It is assumed that inputted speech is processed in the brain through the extraction of acoustic features, phoneme sequence segmentation, and word sequence segmentation. Furthermore, supplementary input is provided by semantic information obtained from other multimodal stimuli.

To identify the ROI involved in DAA, the extent to which a preliminary survey was conducted for each area involved in spoken language processing is represented in Fig.~\ref{fig:LanguageRelatedBrainNetwork}.
The correspondence between the name of each region and its abbreviation is provided in the caption of Fig.~\ref{fig:LanguageRelatedBrainNetwork}.
The main areas of investigation included areas within the ventrolateral prefrontal cortex (VFC), dorsolateral temporal neocortex (DLTC), midlateral temporal cortex (MTC), and inferolateral temporal cortex (ITC).
In addition, the study focused on the anterior temporal lobe (ATL), including A21r, A20r, and temporal polar cortex (A38), because the ``hub-and-spoke'' theory suggests that word meaning is integrated in the ATL~\cite{patterson2007you,rogers2004structure}.
Throughout the present investigation, A38 was connected to the DLTC, which processes language, as well as the ITC, which processes vision, and the caudal division of the posterodorsal (superior) parietal cortex (A7), which processes somatosensory signals. 
The hub was confirmed to have the appropriate anatomical structure.

During the DAA calculation process set up as a TLF, the input is an acoustic signal obtained from the outside world, and the output is a word to which meaning can be mapped. 
Therefore, it would be reasonable to assume that the input is the auditory cortex (A1C) from which the acoustic signal is obtained and the output is A38 to which the word is assigned meaning.
Therefore, in this study, DLTC and A38 in Fig.~\ref{fig:LanguageRelatedBrainNetwork} were set as ROIs.

\subsection{Anatomical and physiological findings}

\begin{figure*}[!tb]
  \begin{center}
    \includegraphics[width=140mm]{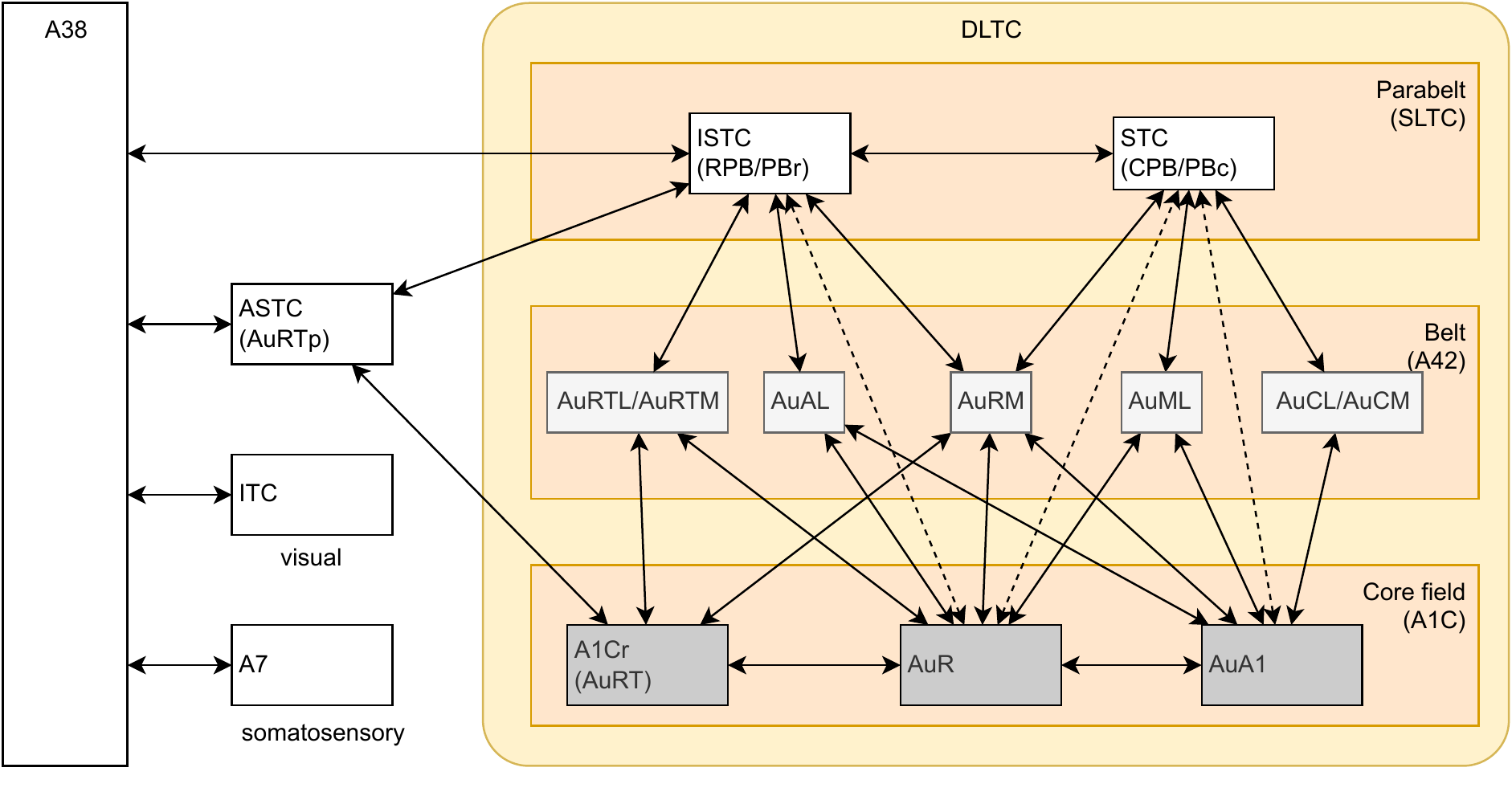}
    \caption{
    Brain information flow (BIF) related to double articulation analysis (DAA). The bi-directional arrows indicate the connections between regions in the cortex.
    }
    \label{fig:BIFforDAA}
  \end{center}
\end{figure*} 

\begin{table*}[!tb]
    \begin{center}
    \caption{
        Assumptions of Non-Human Primate and Human Mapping in the Peripheral Auditory Cortex
    }
    \begin{tabular}{|lp{40mm}|lp{22mm}|lp{50mm}|}
       \hline 
       \multicolumn{2}{|c|}{Non-human primate} & \multicolumn{2}{|c|}{Human [Hamilton, 2001]} & \multicolumn{2}{|c|}{Human [DHBA]} \\	
       ID & Names & ID & Names & ID & Names \\
       \hline \hline
       Core & core field (included AuA1, AuR, RT) & HG & Heschl's gyrus & A1C & primary auditory cortex (core) \\
       \hline
       AuA1 & caudal core field & \multirow{2}{*}{pmHG} & \multirow{2}{*}{posteromedial HG} & \multirow{2}{*}{A41} & \multirow{2}{*}{main portion of A1C (area TC, area 41)} \\
       \cline{1-2}
       AuR & rostal auditory core field &  &  &  &  \\
       \hline
       AuRT & rostral temporal core field & alHG & anterolateral HG & A1Cr & rostral portion of A1C \\
       \hline
       Belt & (include AuRTL, AuRTM, AuAL, AuRM, AuML, AuCL, AuCM) & PT & planum temporale & A42 &	secondary auditory cortex (belt, area 42)\\
       \hline
       Parabelt & (include RPB, CPB) & STG & superior temporal gyrus & SLTC & superior temporal cortex (include ASTC, ISTC, STC) \\
       \hline
       RPB & rostral parabelt & mSTG & middle superior temporal gyrus & ISTC & intermediate superior temporal cortex (area 22i)\\
       \hline
       CPB & caudal parabelt & pSTG & posterior superior temporal gyrus & STC & posterior (caudal) superior temporal cortex (area 22c) \\
       \hline
       AuRTp & rostrotemporal polar belt area & PP & planum polare & ASTC & anterior (rostral) superior temporal cortex (area 22r)\\
       \hline
    \end{tabular}\\
    \vspace{1mm}
    Abbreviations: AuRTL, rostrotemporal-lateral belt; AuRTM, rostrotemporal-medial belt; AuAL, anterolateral belt; \\
    AuRM, rostromedial belt; AuML, middle lateral belt; AuMM, middle medial belt; AuCL, caudolateral belt; AuCM, caudomedial belt.
    \label{tab:CrossSpeciesTable}
    \end{center}
\end{table*}

To apply the SCID method, it is necessary to identify the more detailed anatomical structures of the ROI. Therefore, DLTC and A38, which were established as ROIs, were further investigated.

\subsubsection{Assumptions for mapping cross-spices brain regions in ROI}
Unfortunately, the granularity of regional segmentation for the human auditory cortex is too rough. In contrast, for the non-human primate, a consensus has been reached on the regional segmentation of the auditory cortex as anatomical knowledge has accumulated, starting with early studies~\cite{Kaas2000-ve,Baumann2013-lt}.

Therefore, it was assumed that the anatomical structure of the non-human primate corresponds to that of humans as shown in Table~\ref{tab:CrossSpeciesTable}. The auditory cortex of the non-human primate consists of a core field, which receives frequency-specific intensity vectors of acoustic information from the medial geniculate nucleus (MGN), and a belt field, which surrounds the core field~\cite{Kaas2000-ve,Romanski2009-fg,Scott2017-im,Jasmin2019-wb}.
The core field consists of AuA1, AuR, and AuRT; the belt field consists of AuRTL, AuRTM, AuAL, AuRM, AuML, AuCL, and AuCM. The region on the lateral side of the auditory cortex is called the Parabelt, which contains the RPB and CPB.

The mapping of the brain regions by Hamilton et al.~\cite{Hamilton2021-fc} (specifically referred to in this study) to the DHBA brain regions employed as a standard in the SCID method are shown in Table~\ref{tab:CrossSpeciesTable}.
The core field in non-human primates is called Heschl's gyrus (HG) by \cite{Hamilton2021-fc} and the primary auditory cortex (core), which is A1C in the DHBA. 
AuA1 and AuR inside the Core correspond to pmHG \cite{Hamilton2021-fc} and A41 in the DHBA. AuR inside the Core corresponds to alHG in \cite{Hamilton2021-fc} and A1Cr in the DHBA. The seven subregions in the Belt are mapped to alHG \cite{Hamilton2021-fc} and A42 in the DHBA.
This region is likely to require reconsideration in the future, as it is likely to be affected by the evolution of language in humans~\cite{Marie2018-om}. RPB and CPB in the Parabelt map to mSTG and pSTG \cite{Hamilton2021-fc} and ISTC and STC in the DHBA. The mapping of AuRTp in non-human primates is most controversial. Considering its anatomical connection structure and its location in the brain, it was assumed to map to PP in Hamilton and ASTC in the DHBA in this study.
This study mainly referred to the DHBA~\cite{Ding2017-lc} for humans and the Marmoset information~\cite{Markov2014-ly} for non-human primates.

\subsubsection{Connections}

Figure~\ref{fig:BIFforDAA} shows the BIF constructed as a result of surveys on DAA.
The projection structure between the neocortical areas has feedforward and feedback orientations, which are useful in the design of PGMs.
However, only partial information on the orientation of the projection structure is currently available, and the orientation information in the present study will not be used. For this reason, it was assumed that all connections between brain regions are bidirectional.

The close connections between the various regions contained in the Core and Belt, and the connections between them and the Parabelt, are based on the findings of the non-human primates~\cite{Scott2017-im, Jasmin2019-wb, Kaas2000-ve}.
Connections between the various regions in the Parabelt were determined using human~\cite{Fan2016-wp} and Marmoset connectome data~\cite{ Markov2014-ly}.

For the temporal pole (A38), in addition to the connection to the DLTC involved in spoken language~\cite{Scott2017-im,Jasmin2019-wb,Markov2014-ly,Fan2016-wp}, connections to the inferior temporal cortex (ITC)~\cite{Pascual2015-ju} and the somatosensory region, A7~\cite{Fan2016-wp}, have been identified.

\subsubsection{Physiological findings}

Through physiological studies, it has been shown that the computational function performed by the DLTC in the language-related brain network, Fig.~\ref{fig:LanguageRelatedBrainNetwork}, is as follows.
Tonotopic acoustic processing in the auditory cortex (A1C, A42)~\cite{Kayser2007-ee}.
In STC and ISTC, language-specific phonetic features (voiced, velar, plosive, etc.), relative pitch change features, and temporal landmarks, among others, are processed~\cite{Hamilton2021-fc,Bhaya-Grossman2022-gg}.

\subsection{Functions}
This section introduces functions of the areas involved in speech processing.

\subsubsection{Speech language processing}
A1C is responsible for the input of auditory stimuli, while the DLTC is thought to be responsible for the processing of nonverbal concepts, phonological storage, and the formation of phonological representations~\cite{minagawa2017infant}.
A39 and A40 are thought to store phonological and articulatory relationships, while the PMC is thought to be involved in speech perception and mouth movement recognition, as well as phonological storage and retrieval~\cite{minagawa2017infant}.
The VFC is believed to be responsible for phonological storage and retrieval as well as speech category discrimination~\cite{minagawa2017infant,vigneau2006meta}.
It is reported that the superior temporal cortex and superior temporal sulcus perform the spectral time series and phonological analyses. 
It is also believed that the anterior temporal cortex and the MTC, containing A21r, A21c, and A37, are involved in speech recognition and the representation of lexical concepts~\cite{chang2015contemporary}.
The process of passing through the parietotemporal border, PMC, and VFC regions is thought to perform sensorimotor integration by mapping phonological features to articulomotor representations~\cite{chang2015contemporary}.
There are two dorsal pathways, one leading to A44 and the other to the PMC, each of which is believed to involve separate auditory-motor interactions~\cite{chang2015contemporary}.
One is associated with individual segmented speech sounds and is used to acquire and retain the basic articulatory skills of speech sounds.
The other pathway, involving sequences of segmented speech sounds, is said to enable the learning of new lexicons.

The area around the STC is responsible for recitation and phonology, A38 is responsible for the assignment of semantic information, and the VFC (especially the area corresponding to A44 in the operculum) is responsible for syntax~\cite{friederici2011brain}.
The input auditory stimuli go to the STC for phonological storage and retrieval.
In addition, the visual stimuli are integrated and assigned meaning at A38 to form a sentence in the VFC (A44, A45).
The process from A38 to A44 supports higher-order linguistic processes of local phrase structure construction based on word category and the combination of adjacent elements in the flow~\cite{friederici2011brain}.
Therefore, the region that is thought to be responsible for word segmentation is considered to be A38.
Assuming this is the case, it is important to know the role of each part and the type of information processes in each pathway.


\subsubsection{Multimodality integration}

The representation of semantic cognition was proposed as a ``hub-and-spoke'' theory of semantic representation~\cite{patterson2007you,rogers2004structure}.
Semantic cognition refers to the ability to use and manipulate knowledge acquired throughout life to generalize and support verbal/nonverbal behaviors.
Modality-specific information sources (spokes) are coded across a set of processing units in separate processing layers within the model.
Each spoke layer is interconnected to a single transmodal hub.
For example, the model is given the visual shape of each item as input and trained to reproduce the sounds, names, values, and other information associated with each item.
As a result of this training, the model forms a generalizable semantic representation~\cite{benavides2021variation,bonner2013anterior}.

The hubs that pass this semantic information are located within the ATL regions, and the modality-specific spokes are distributed in different neocortical regions.
Each spoke passes information bi-directionally to and from the ATL hub through short- and long-range white matter coupling pathways.
In particular, A38 is considered to handle multimodal information because of the connections to visual and somatosensory stimuli input pathway~\cite{ralph2017neural,sierpowska2019white,bajada2015transport}.
This region may be also used in learning words.

\section{Brain-inspired Probabilistic Generative Model for Double Articulation Analysis}
In this section, a novel computational model is proposed for DAA as a brain reference architecture (BRA).
The brain-inspired PGM for DAA from speech input is constructed based on the functional hypothesis derived from neuroscientific findings and existing PGMs.

\subsection{Functional hypotheses from neuroscientific findings}
%
%
We have a hypothesis that areas from A1C to A38 may drive DAA in the brain. 
The respective functions of A1C, A42, SLTC, and A38 are assumed to be the sound input, the discrimination of speech/non-speech and speech perception in a phoneme sequence, the syllabification of speech signals, and the word segmentation, respectively.
From there, it is thought that segmentation into words is performed at A38, and complex sentences are constructed using word category at A44.
The knowledge that the ASTC processes the acoustic characteristics suggests that the ASTC is used for the categorical representation of auditory features.
Furthermore, it is conceivable that integrated multimodal information based on hub-and-spoke theory plays an auxiliary role in word segmentation in A38.

\begin{figure*}[tb]
  \begin{center}
    \includegraphics[width=1.0\linewidth]{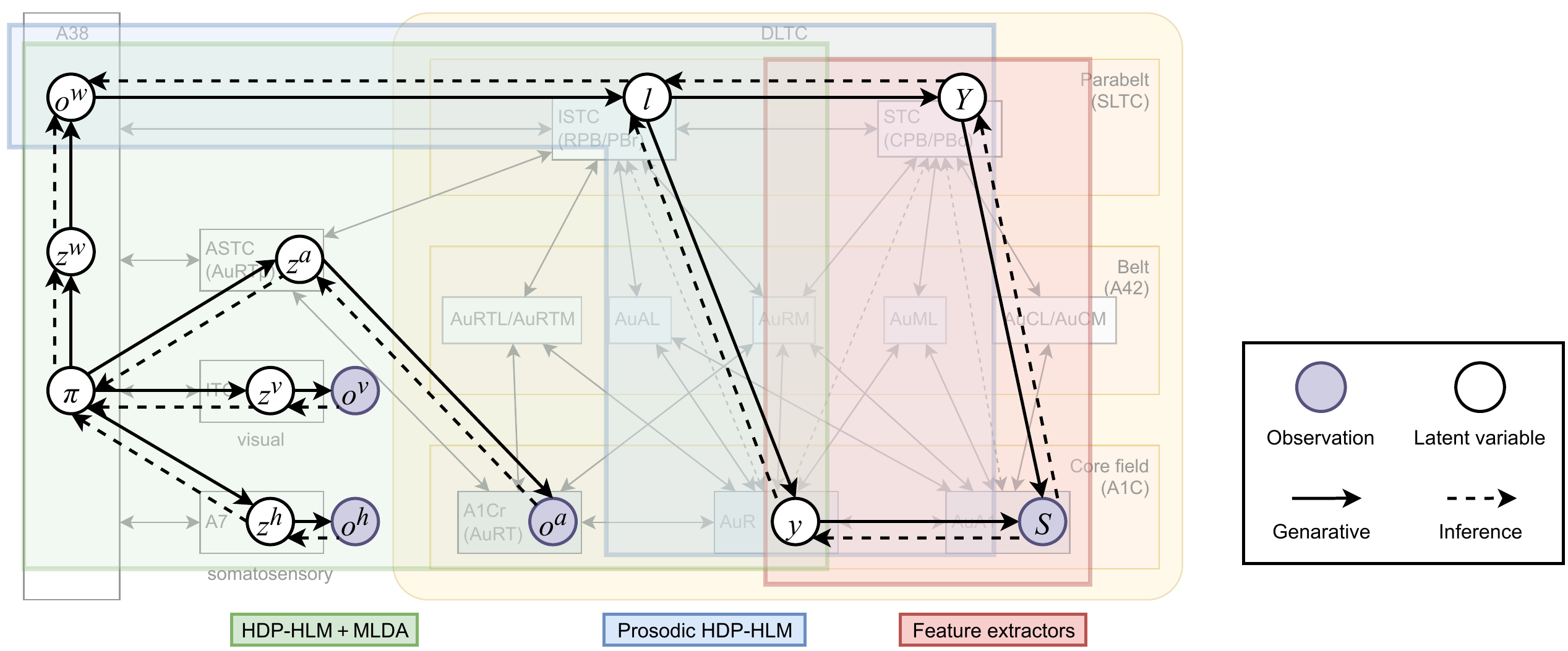}
    \caption{
    Graphical model representation of the probabilistic generative model (PGM) with a brain information flow (BIF) for double articulation analysis (DAA) in the brain.
    The black directed and dotted arrows indicate the direction of the generative and inference processes for DAA in the PGM, respectively.
    Note that global model parameters, hyperparameters, and subscripts representing time steps are omitted from the description.
    }
    \label{fig:apr_bif_pgm}
  \end{center}
\end{figure*} 

\begin{table}[tb]
    \begin{center}
    \caption{
        Description of variables in the PGM for DAA
    }
    \begin{tabular}{ccp{160pt}} \hline
        \textbf{Region} & \textbf{Symbol} & \textbf{Function of components on HCD} \\ \hline
        AuA1 & $S$     & Spectrum Information about speech signal \\ 
        AuR  & $y$     & Acoustic features (phonological information) about speech signal, e.g., MFCC \\
        STC  & $Y$     & Prosodic feature about speech signal, e.g., fundamental frequency and silent interval \\         
        ISTC & $l$     & Phoneme sequence (including representations of the index and duration length of phonemes)\\ 
        A38  & $o^{w}$ & Segmented word sequence \\ 
        A38  & $z^{w}$ & Latent variable representing the category assigned to each word \\ 
        A38  & $\pi$   & Integrated object category information representing the probability of occurrence of each category, i.e., hub\\         
        A1Cr & $o^{a}$ & Auditory feature about an object \\ 
        ASTC & $z^{a}$ & Latent variable representing the category assigned to each feature in auditory modality  \\         
        A7$^\ast$  & $o^{h}$     & Somatosensory features of an object \\ 
        A7$^\ast$   & $z^{h}$     & Latent variable representing the category assigned to each feature in somatosensory modality \\ 
        ITC$^\ast$  & $o^{v}$     & Visual features of an object \\ 
        ITC$^\ast$  & $z^{v}$     & Latent variable representing the category assigned to each feature in visual modality \\ 
        \hline
    \end{tabular}\\
    \vspace{1mm}
    $\ast$ This functional description omits detailed processing on the pathway.
    \label{tab:graphical-model}
    \end{center}
\end{table}

\subsection{Probabilistic generative model}
The proposed model of this study is shown in Fig.~\ref{fig:apr_bif_pgm}, and the definitions of variables are shown in Table~\ref{tab:graphical-model}.
The strength of the proposed model is that its feasibility is supported by the fact that the existing models partially realize DAA.
The proposed model was created as an integrated model of two existing PGMs: Prosodic HDP-HLM~\cite{okuda2021double} and HDP-HLM+MLDA~\cite{Taniguchi2022UnsupervisedMW}.
Thus, this model takes advantage of all three types of cues in the lexical acquisition reported by Saffran et al.~\cite{saffran1996word}.
In addition, while these existing models assume the generative process of auditory features $y$ and $Y$, the proposed model assumes the spectrum of the speech signal $S$ to observation as a source stimulus.
Therefore, the brain-inspired process of speech feature extraction is also incorporated in the PGM.
As a detail of the aforementioned contents with references, a pre-screening version of the BRA data (``DAA.bra'') was released on the site of the Whole Brain Architecture Initiative\footnote{https://docs.google.com/spreadsheets/d/1nINeM7MFcTqPyQJGqPDZp1e\\mMwgkTpUi0OXoeJvgUqI/}.

%
As an inference process of Fig.~\ref{fig:apr_bif_pgm}, first, a speech stimulus $S$ is input.
At this time, non-speech auditory features $o^{a}$, somatosensory features $o^{h}$, and visual features $o^{v}$ enter as co-occurrence cues related to an object.
Next, the speech stimulus $S$ is distinguished into acoustic features $y$ and prosodic features $Y$.
Based on the two features, a phoneme sequence $l$ is obtained, and further transformed into a word sequence $o^{w}$.
Here, the multimodal object category information obtained from $\pi$ provides cues for word segmentation via the latent variable $z^{w}$.
$\pi$ integrates the multimodal object category information, including $z^{w}$, $z^{a}$, $z^{h}$, and $z^{v}$, which are latent variables representing the category assigned to each feature.
From the above, this graphical model represents the DAA's function of generating a phoneme sequence $l$ and a word sequence $o^{w}$ using speech $S$ as the main input, while integrating other modality features $o^{a}$, $o^{h}$, and $o^{v}$.

Each latent variable is associated with an acoustic model for $y$, a prosody model for $Y$, a word model (letter bi-gram and word dictionary) and duration distribution for $l$, and a language model (word bi-gram) for $o^{w}$ as global variables.
The phoneme sequence $l$ contains representations of the index and duration of phonemes.
Markovianity is assumed for the word sequence $o^{w}$ and the phoneme sequence $l$ with respect to words and phonemes, respectively.

\subsection{Consistency of model with scientific knowledge}

%
%
%
Each node of the PGM can be superimposed on a region on the BIF based on its function, as shown in Fig.~\ref{fig:apr_bif_pgm}.
%
A1C is the sound input region where not only speech $S$ but also general sound $o^{a}$ including environmental sound are received.
%
Speech stimuli $S$ extract acoustic features $y$ in A1C.
Also, $S$ are sent to A42, which distinguishes between speech and non-speech and identifies a prosodic feature $Y$ in the STC.
The input is converted to a phoneme sequence $l$ in the ISTC.
It is consistent with findings such as learning hypersegmental sounds, which are intonational and rhythmic features of the native language, and learning phoneme-level segmental sounds~\cite{homae2006right}.
In addition, the phoneme sequence $l$ is sent from the ISTC to A38, where it becomes the word sequence $o^{w}$.
The HDP-HLM~\cite{taniguchi2016nonparametric} assumes n-grams for word transitions, which is consistent with the function of A38.
A38 belonging to the ATL is also consistent with the existence of $\pi$ from the hub-and-spoke theory. 
A38 is also connected to A7 and the ITC.
From the above, the assignment between regions in the BIF and variables in the PGM are supported.

In the Core/Belt section, which is A1C/A42, general acoustic recognition is processed. 
The speech spectrum $S$ was placed in AuA1 because A1C are the main receivers of information. 
In addition, $y$ was placed in AuR from tonotopic sound processing in the primary auditory cortex~\cite{Hamilton2021-fc}.
A42 and the SLTC are known to perform human-limited speech language-dependent processing.
Prosodic features are placed in the STC~\cite{Hamilton2021-fc,Bhaya-Grossman2022-gg}. 
Given that word segmentation occurs in A38, it would be more appropriate for the phoneme sequence $l$ to be represented in the ISTC and $Y$ placed in the STC.
Acoustic processing is rapid, and the period increases as one goes from AuA1 and AuR to the SLTC, which can be accommodated in a semi-Markov model.
Therefore, the argument of the \cite{Hamilton2021-fc} is consistent with our proposed engineering treatment.

Our proposed PGM states that it is a process in the superior temporal sulcus and the SLTC, similar to \cite{Bhaya-Grossman2022-gg}.
In addition, it is modeled in a more modularized manner than the study \cite{Bhaya-Grossman2022-gg}. 
The anatomical structure used in this case is the parallel pathway~\cite{Jasmin2019-wb}. 
In the proposed model, the distinction between $y$ and $Y$ corresponds to the claim of paralleled processing \cite{Hamilton2021-fc}.
Thus, from an evolutionary perspective, it seems reasonable to consider two processing paths in speech feature extraction.

\section{Conclusion}
\label{sec:conclusion}

In this paper, a PGM for DAA of spoken language was constructed based on a method for developing a BRA. 
The anatomical structures and functions of the brain involved in spoken language processing were investigated and organized as BRA data. 
The results showed that the existing computational models of DAA is consistent with neuroscientific findings.
The engineering feasibility of the proposed model is supported by existing models.
We believe that the proposed model will have higher performance than the existing models.


As a future task, the proposed model will be implemented.
In addition, the survey content suggests further model extensibility for the BIF areas that are not covered by the proposed PGM. 
This study mainly investigated the left hemisphere, but functional localization~\cite{homae2006right} and functional assistance by both hemispheres~\cite{friederici2011brain} for auditory and language processing are also issues to be considered.






\balance


\bibliographystyle{IEEEtran}
\bibliography{IEEEabrv,bibliography,articles} 

\end{document}